\documentclass[prl,aps,showpacs,twocolumn]{revtex4}

\makeatletter
\def\@pnumwidth{2em}
\makeatother
\usepackage{graphicx}
\usepackage{dcolumn}
\usepackage{bm}

\begin{document}
\title{Family of Solvable Generalized Random Matrix Ensembles with Unitary Symmetry}
\author{K. A. Muttalib and J. R. Klauder}
\thanks{Also Department of Mathematics.}
\affiliation{Department of Physics, University of
Florida, P.O. Box 118440, Gainesville, FL 32611-8440}

\begin{abstract}

We construct a very general family of characteristic functions
describing Random Matrix Ensembles (RME) having a global unitary
invariance, and containing an arbitrary, one-variable probability
measure which we characterize by a `spread function'. Various
choices of the spread function lead to a variety of possible
generalized RMEs, which show deviations from the well-known
Gaussian RME originally proposed by Wigner. We obtain the
correlation functions of such generalized ensembles exactly, and
show examples of how particular choices of the spread function can
describe ensembles with arbitrary eigenvalue densities as well as
critical ensembles with multifractality.

\end{abstract}


\maketitle

The concept of Random Matrix Ensembles (RME), originally proposed
by Wigner to describe statistical properties of the eigenvalues
and eigenfunctions of complex nuclei \cite{Mehta}, have proved to
be a very useful idea in the studies of a wide variety of physical
systems including equilibrium and transport properties of
disordered quantum systems, quantum chaos, two-dimensional quantum
gravity, conformal field theory and chiral phase transitions in
quantum chromodynamics as well as financial correlations and
wireless communications \cite{reviews,fin}. The underlying reason
for such a wide range of applications is the universality of the
correlations between eigenvalues of given classes of RMEs. For
example, once appropriate variables are chosen in which the mean
spacing between eigenvalues is unity, the Gaussian ensembles of
random $N\times N$ matrices (with given symmetries) have, in the
$N\rightarrow\infty$ limit, a universal (zero parameter) form for
the nearest neighbor spacing distribution or the spectral rigidity
(number variance of eigenvalues in a given range) known generally
as the Wigner-Dyson distribution \cite{Mehta}. The appropriately
scaled energy levels of complex nuclei and transmission levels of
weakly disordered mesoscopic metals both follow the above
universal distributions even though the physical sizes of the
systems differ by about nine orders of magnitude \cite{smmp}.

More recently, much interest has been generated in finding RMEs
that \textit{deviate} from the universal properties of the
Gaussian Ensembles in specific ways. One particular example is the
attempt to find `critical' ensembles relevant for systems at the
critical point of e.g.~the Anderson transition in disordered
conductors where the spacing distribution or the spectral rigidity
is known to deviate from those of Gaussian RMEs
\cite{critical,critical2,mut}. Another example is the attempt to
find a one-parameter generalization of the Gaussian RME that can
describe a monotonic change from the universal Wigner-Dyson
distribution to a completely uncorrelated Poisson distribution as
the parameter is changed, which may be relevant for a crossover
from a chaotic to an integrable system \cite{crossover}. Yet
another example is the observation for financial cross-correlation
matrices in which statistics of most eigenvalues agree with the
universal predictions of Gaussian RMEs but there are deviations
for a few of the largest eigenvalues \cite{stan}. It is therefore
of great interest to a wide variety of areas and disciplines to
study RMEs that are in some sense generalizations of the Gaussian
RMEs.

In seeking generalizations of Gaussian RMEs, suitable for
arbitrarily many variables, it is natural to begin with the
probability distributions $P_N(X)$, where $X$ denotes an $N\times
N$ Hermitian matrix. Gaussian ensembles have centered
distributions that are in fact (exponential) functions of the
single variable $tr(X^2)$. It is natural to restrict our
generalizations to probability distributions
$P_N(X)=W_N(tr(X^2))$. One such example that might come to mind
could be $P_N(X)\propto \exp (-[tr(X^2)]^2)$, but although such a
proposal is satisfactory for any \textit{finite} $N$, it fails to
generalize to a valid new distribution as $N\rightarrow\infty$.
The clue to discovering the proper class of generalizations is to
work not directly with the distribution $P_N(X)$ themselves, but
with their Fourier transforms, i.e., with the associated
\textit{characteristic functions} \cite{book} $C_N(T)$ given by
\begin{equation}\label{cf}
C_N(T)=\int e^{i~ tr (TX)}P_N(X)dV_X
\end{equation}
with $C_N(0)=1$. The integration is over the invariant Haar
measure that preserves hermiticity. In the present work we first
prove that if $C_N(T)$ is a function of $tr (T^2)$ only, then the
\textit{most general} $C_N(T)$, valid for arbitrarily large $N$,
can always be written as
\begin{equation}\label{C}
C_N(T)=\int_0^{\infty}e^{-b~tr (T^2)}f(b)db\; , \;\;\;
\int_0^{\infty}f(b)db=1
\end{equation}
where $f(b)$ is \textit{any} non-negative function, which may be
chosen phenomenologically. While Eq.~(\ref{C}) is evidently a
positive superposition of Gaussians, a vast family of
distributions is thereby included, e.g., Cauchy, Levy, etc.; this
result generalizes the specialized and more restrictive examples
of [\onlinecite{bertoula}]. We then show that the $n$-point
correlation function for the corresponding unitary RME \cite{note}
of the matrices $X$ with eigenvalues $x_i$ can be written down
\textit{exactly} as
\begin{equation}\label{Rn}
R_n(x_1 \cdots x_n) = \int_0^{\infty} \frac{db f(b)}{(4b)^{n/2}}
det[K^G_N(\bar{x}_i,\bar{x}_j)]_{i,j=1,2, \cdots n}.
\end{equation}
Here we have defined $\bar{x}_i=x_i/2\sqrt{b}$, and
$K^G_N(x_i,x_j)=\sum_{n=0}^{N-1}\varphi_n(x_i)\varphi_n(x_j)$ is
the well-known 2-point kernel of the Gaussian RME \cite{Mehta},
where $\varphi_n(x)\propto e^{-x^2/2}H_n(x)$ are orthonormal
functions associated with the Hermite polynomials $H_n(x)$. In
particular, the one-point function is just the density of levels,
given (in the large $N$ limit) by
\begin{equation}\label{sig}
\sigma_N(x)=
\frac{\sqrt{2N}}{2\pi}\int_{x^2/8N}^{\infty}\frac{dbf(b)}{b^{1/2}}
\sqrt{1-x^2/8Nb},
\end{equation}
where we have used the known result that for large $N$,
$K^G_N(x,x)\equiv\sigma^G_N(x)=\sqrt{2N-x^2}/\pi$ for $|x|<
\sqrt{2N}$ and zero otherwise \cite{Mehta}. It is clear that one
can obtain different $x$ and $N$ dependence for the densities by
choosing different $f(b)$. Similarly, the two point cluster
function defined as $T_2(x_1,x_2)\equiv -R_2(x_1,
x_2)+R_1(x_1)R_1(x_2)$ has the form $T_2=T^0_2-\delta T_2$, where
\begin{equation}\label{T2}
T^0_2  (x_1,x_2)  =  \int_0^{\infty}\frac{dbf(b)}{4b}
[K^G_N(\bar{x}_1,\bar{x}_2]^2
\end{equation}
and
\begin{equation}\label{dT}
\delta T_2  = \int_0^{\infty} \frac{dbf(b)}{4b}
\sigma^G_N(\bar{x}_1)\sigma^G_N(\bar{x}_2)-
\sigma_N(x_1)\sigma_N(x_2).
\end{equation}
We call $f(b)$ the `spread function'. Note that the Gaussian RME
corresponds to the choice $f(b)=\delta(b-b_0)$, for which $\delta
T_2$ is identically zero. Other choices of the spread function can
describe a variety of possible generalized RMEs. Alternatively,
correlation functions of a physically relevant RME of matrices $X$
characterized by a given $C_N(T)$ can be obtained exactly if the
corresponding spread function $f(b)$ can be identified. For
example, $C_N(T)= e^{-b_0 \sqrt{tr~T^2}}$ will correspond to a
choice of $f(b)\propto b^{-3/2}e^{-b_0^2/b}$, for which the exact
correlation functions can be easily written down.

The universal features of Gaussian RMEs arise in the
$N\rightarrow\infty$ limit and when the variables are chosen in
which the mean level spacing is unity (this is known as
`unfolding'). In place of $x_1$ and $x_2$ one defines new
variables $\rho$ and $\zeta$ such that $d\rho=\sigma(x_1)dx_1$,
$d\zeta=\sigma(x_2)dx_2$, and the new cluster function
\begin{equation}\label{Y}
Y_2(\rho,\zeta)d\rho d\zeta\equiv T_2(x_1,x_2)dx_1dx_2
\end{equation}
is well defined everywhere in the limit $N\rightarrow\infty$. For
Gaussian RMEs, there exists the sum rule that the integral of
$Y_2(\rho,\zeta)$ over $\rho$ is always unity. It has been argued
\cite{ckl,can} that the violation of this sum rule is a signature
of critical ensembles, where the deficit of the sum rule (for
translationally invariant cluster functions)
\begin{equation}\label{eta}
\eta\equiv 1-\int_{-\infty}^{\infty}Y_2(\rho,\zeta)d\rho
\end{equation}
is related to the multi-fractality of wave functions at the
critical point \cite{mf}. We will show that there are choices for
$f(b)$ for which the sum rule is violated. These choices would
then correspond to critical ensembles.

We begin by proving that if $C(T)$ is a function of $tr~ (T^2)$,
then the most general $C_N(T)$, valid for arbitrarily large $N$,
can be written in the form Eq.~(\ref{C}). Note that if $C(T)$ is a
function of $tr~ (T^2)$ only, then from Eq.~(\ref{cf}) we have
$P_N(U^{\dag}XU)=P_N(X)$ where $U$ is unitary. Thus the
distribution is invariant under a rotation of basis.

\textit{Proof of Eq.~(\ref{C})}: We suppose that
\begin{equation}\label{K-1}
C(T)\equiv E(\|T\|^2_2)\equiv \int e^{i\,tr(TX)}\,d\mu(X)\;,
\end{equation}
for all Hermitian $T$ with $\|T\|^2_2\equiv tr(T^2)<\infty$, where
$\mu$ is a suitable probability measure. It  follows that
$C(T-S)=E(\|T-S\|^2_2)$ is a real, continuous function of positive
type, and the GNS Theorem \cite{GNS} ensures that there exist
vectors $|T\rangle$ and $|S\rangle$ in a separable Hilbert space
such that
\begin{equation}
\langle S|T\rangle\equiv E(\|T-S\|^2_2)\;.
\end{equation}
A family of operators $V(U)$ may be defined by
\begin{equation}
\langle S|V(U)|T\rangle\equiv \langle
S|T+U\rangle=E(\|T+U-S\|^2_2)\;,
\end{equation}
and it readily follows that $V(U)$ is an Abelian group of unitary
operators for which the operator norm $\|V(U)\|=1$. Now consider
the sequence $U^{(M)}=\{U^{(M)}_{rs}\}$ where
\begin{equation}
U^{(M)}_{rs}\equiv \sqrt{u}\; \delta_{rM}\delta_{sM}\;,\hskip.3cm
u\ge0\;,\hskip.3cm 1\le M<\infty\;.
\end{equation}
It follows that the weak operator limit
\begin{equation}
\lim_{M\rightarrow\infty}\langle
S|V(U^{(M)})|T\rangle=E(\|T-S\|^2_2+u)\equiv\langle
S|A(u)|T\rangle\;,
\end{equation}
an expression that defines the operator $A(u)$ for all $u\ge0$.
Clearly, $A(u)^\dagger=A(u)$, $A(u)A(v)=A(u+v)$, and $\|A(u)\|\le
1$, and therefore $A(u)=e^{-uB}$, where $B^\dagger=B\ge0$. Hence,
\begin{equation}
E(u)=\langle 0|e^{-uB}|0\rangle=\int_0^\infty e^{-ub}\,dm(b)\;,
\end{equation}
where $m$ is a probability measure, and the latter relation
follows from the spectral representation for $B$. Finally, if we
assume that $m$ is absolutely continuous and replace $u$ by
$\|T\|^2_2$, we recover Eq.~(\ref{C}). This completes the proof of
Eq.~(\ref{C}).

We next show that given Eq.~(\ref{C}), the $n$-point correlation
function is given by Eq.~(\ref{Rn}).

\textit{Proof of Eq.~(\ref{Rn})}: Given the characteristic
function Eq.~\ref{C}, the probability density is, from
Eq.~\ref{cf},
\begin{equation}
P_N(X) \propto \int_0^{\infty} db f(b) \int e^{-i~ tr (TX)}e^{-b~
tr (T^2)} dV_T .
\end{equation}
The integral over $N^2$ independent elements of $T$ results in
\begin{equation}
P_N(X) \propto \int_0^{\infty}\frac{db f(b)}{b^{N^2/2}}   e^{- tr
(X^2)/4b}.
\end{equation}
The joint probability distribution of the eigenvalues of $X$ is
then given by \cite{Mehta}
\begin{equation}
P_N(\{x_i\}) \propto \int_0^{\infty}\frac{db f(b)}{b^{N^2/2}}
\prod_{i<j}(x_i-x_j)^2 e^{- \sum_i x^2_i/4b}\; ,
\end{equation}
where $\prod_{i<j}(x_i-x_j)^2$ is just the Jacobian of
transformation from the matrix element to the
eigenvalue/eigenvector coordinates. We now take advantage of the
known results for Gaussian Random Matrix Ensembles by noting that
before the $b$-integral, a change of variables
$\bar{x}_i=x_i/2\sqrt{b}$ changes the distribution to exactly
Gaussian RME with some additional $b$-dependent terms; this leads
directly to Eq.~(\ref{Rn}). This completes the proof of
Eq.~(\ref{Rn}).

As an example, let us consider the spread function $f(b)\propto
b^{N^2/2+\nu-1}e^{-\varepsilon^2/4b}e^{-\gamma b}$. Then
\begin{equation}
P_N(X) \propto
\left(\frac{\varepsilon^2+tr~(X^2)}{4\gamma}\right)^{\nu /2}
K_{\nu}\left(\sqrt{\gamma (\varepsilon^2+tr~(X^2))}\right)\; ,
\end{equation}
where $K_{\nu}$ is a modified Bessel function. In the limit
$\gamma\rightarrow 0$, choices of $\nu$ would include ensembles of
L\'{e}vy matrices \cite{levy}. If $\nu=-n$ where $n$ is a positive
integer, then in the same limit $\gamma\rightarrow 0$ this gives
rise to $P_N(X)\propto (\varepsilon^2+tr~(X^2))^{-n}$, while in
the opposite limit $\gamma \rightarrow \infty$ we get
$P_N(X)\propto
(\varepsilon^2+tr~(X^2))^{-(n/2+1/4)}e^{-\sqrt{\gamma
(\varepsilon^2+tr~(X^2))}}$. The 2-point correlation function for
all $\gamma$ and $\varepsilon$ can be written down exactly from
Eqs.~(\ref{T2}) and (\ref{dT}).

As an explicit example of how novel behavior of the correlation
functions may arise for some choices of $f(b)$ in the large $N$
limit, let us consider the level density for
\begin{eqnarray}
f(b) &=& c_N\frac{2\bar{b}+1}{[\bar{b}(\bar{b}+1)]^{3/2}}, \;\;\;
b<b_0;\cr &=& 0, \;\;\; b>b_0
\end{eqnarray}
where $\bar{b}\equiv\sqrt{8Nb}$ and $b_0$ is determined from the
normalization condition. We choose $c_N$ such that
$\bar{b}_0\equiv \sqrt{8Nb_0} \gg 1$. Then the density obtained
from Eq.~(\ref{sig}) is given by
\begin{equation}
\sigma_N(x)\approx\frac{1}{\pi}\frac{c_N}{\sqrt{x(x+1)}}\;,
\end{equation}
which is similar to that satisfied by the transmission eigenvalues
in disordered conductors and is known to lead to deviations from
Gaussian RME \cite{smmp,mut}. Note that the density diverges as
$1/\sqrt{x}$ in the limit $x\rightarrow 0$. It is clear that by
appropriately choosing the spread function $f(b)$, one can obtain
a variety of densities that can go to zero, a constant or infinity
as a function of $x$ as $x\rightarrow 0$. One can also choose
$f(b)$ to obtain an $N$-dependent density at the origin that goes
to zero, a constant or infinity in the $N\rightarrow\infty$ limit.
An $N$-independent finite density at the origin was conjectured to
be important for the novel properties of the $q$-Random Matrix
Ensembles reviewed in \cite{qrme}.

In particular, $f(b)$ can be defined such that in the limit
$N\rightarrow\infty$ both $\sigma(x)$ and $\delta T_2(x,y)$ become
constant.  In such cases if we define
\begin{equation}
\phi(\lambda)= \int_{0}^{\infty} db  f(b)e^{-\lambda/\sqrt{b}}\; ,
\end{equation}
then $\phi(0)=1$,
$\sigma(x)=\sigma_0=-\frac{\sqrt{2N}}{2\pi}\phi'(0)$ and $\delta
T_2(x,y)=\frac{N}{2\pi^2}[\phi''(0)-(\phi'(0))^2]$. We now choose
$\rho=\sigma_0 x$, $\zeta=\sigma_0 y$ to scale the density to
unity. Then the scaled cluster function defined in Eq.~(\ref{Y})
has the simple form $Y_2(\rho,\zeta)=T_2(x_1,x_2)/\sigma^2_0$.
Using Eqs.~(\ref{T2}) and (\ref{dT}) to define $Y^0_2$ and $\delta
Y_2$, we find
\begin{equation}\label{Y2}
Y^0_2(\rho,\zeta)  =  \frac{1}{\pi^2}  \int_0^{\infty} \frac{db
f(b)}{(\rho-\zeta)^2}\sin^2
\left(\sqrt{\frac{N}{2b}}\frac{(\rho-\zeta)}{\sigma_0}\right),
\end{equation}
where we have used the known large $N$ behavior for $K^G_N(x,y)$
\cite{Mehta}, and
\begin{equation}\label{dY}
\delta Y_2 (\rho,\zeta) =  \frac{\phi''(0)}{(\phi'(0))^2}-1\; .
\end{equation}
The deficit of the sum rule Eq.~(\ref{eta}) then takes the form
\begin{equation}\label{eta1}
\eta=1- \int_{-N/2}^{N/2}d\rho Y^0_2(\rho,\zeta)+ N\left[
\frac{\phi''(0)}{(\phi'(0))^2}-1\right].
\end{equation}

If $\sigma_0\propto \sqrt{N}$, the argument of the sine function
in $Y^0_2$ is independent of $N$. Then the integral of $Y^0_2$ is
just unity for all $f(b)$ in the limit $N\rightarrow\infty$. Using
the inequality
\begin{equation}
\int db f(b)b^{-1/2} \leq \sqrt{\int dbf(b)b^{-1}\int db f(b)}
\end{equation}
and the normalization of $f(b)$, we obtain $\eta\geq 0$. The
equality sign holds only if $f(b)\propto \delta(b-b_0)$, which is
the Gaussian RME. As an explicit example of finite positive $\eta$
in the large $N$ limit, let us choose
\begin{equation}\label{fb}
f(b)\propto b^{\alpha}e^{-\beta/\sqrt{b}}e^{-\gamma \sqrt{b}}.
\end{equation}
With $\alpha=-3/4$, this immediately leads to
$\eta=N/2\sqrt{\beta\gamma}$. If $\sqrt{\beta\gamma}\propto N$, we
get a well defined critical ensemble in the thermodynamic limit.
Note that for finite $\eta$, Eq.~(\ref{dY}) gives $\delta
Y_2=\eta/N\rightarrow 0$ in the $N\rightarrow\infty$ limit.

The variance $\Sigma(s)=\langle n^2\rangle-\langle n\rangle^2$ of
the number of eigenvalues $n$ in an interval $(-s/2,s/2)$ is given
by
\begin{equation}
\Sigma(s) = \int_{-s/2}^{s/2}d\rho
\int_{-s/2}^{s/2}d\zeta[\delta(\rho-\zeta)-Y_2(\rho-\zeta)].
\end{equation}
For Gaussian RME, $\Sigma(s)\propto \ln s$ for large $s$. For
critical ensembles discussed in the literature
\cite{critical,mut}, $\Sigma(s)\propto s$. Our choice of the
spread function  $f(b)$ in Eq.~(\ref{fb}), which gives rise to a
constant $\delta Y_2$, produces a term $\Sigma(s)\propto s^2$ in
addition to terms $\Sigma(s)\propto \ln s$ from $Y^0_2$. However,
as mentioned earlier, this term is proportional to $\eta/N$, which
vanishes in the large $N$ limit for finite $\eta$.  Eq.~(\ref{fb})
therefore corresponds to a novel kind of critical ensemble with
$\Sigma(s)\propto \ln s$ for large $s$.

Note from Eq.~(\ref{T2}) that in general $\delta T_2$ is not
translationally invariant; clearly the critical ensembles in such
cases will not give rise to number variance $\Sigma(s)\propto \ln
s$, and will correspond to a different class. We also get a
different class of critical ensembles if the contribution to
$\eta$ from $Y^0_2$ is different from unity. For choices of $f(b)$
which diverge as $b\rightarrow 0$, changing the order of the
integrals over $\rho$ and $b$ in Eq.~(\ref{eta}) may not be
justified, and the integral in Eq.~(\ref{Y2}) may depend on the
choice of such $f(b)$.

In summary, we have constructed a generalized Random Matrix
Ensemble whose characteristic function contains an arbitrary
non-negative spread function $f(b)$ with the only condition that
$\int_0^{\infty}f(b)db=1$. The correlation functions of the
generalized ensembles are exactly solvable for any given $f(b)$.
Various choices of $f(b)$ lead to a variety of possible density of
levels $\sigma_N(x)$, which can depend on $x$ or $N$ in a variety
of different ways, leading to possible deviations from Gaussian
Ensembles. In particular, we showed that it is possible to choose
forms of $f(b)$ that lead to violations of the sum rule for the
scaled 2-point cluster function $Y_2(\rho,\zeta)$ where the
deficit of the sum rule $\eta$, as given in Eq.~(\ref{eta}), is a
characteristic of critical ensembles with multifractal wave
functions. Unlike critical ensembles discussed in the literature,
these sum rule violations can correspond to different forms for
the number variance $\Sigma(s)$, corresponding to different
classes of critical ensembles. These solvable generalized
ensembles should therefore be of interest in a wide range of areas
where Random Matrix Ensembles play an important role. While there
are only a few known examples of generalized RMEs for which
correlation functions can be evaluated exactly
\cite{critical2,mut}, a given model of $C_N(T)$ or $P_N(X)$
characterizing a physically relevant generalized RME becomes
exactly solvable if the corresponding spread function $f(b)$ can
be found. This opens up the possibility to obtain exact results
for a variety of interesting and physically useful generalized
RMEs.


\begin{thebibliography}{40}
\bibitem{note} For simplicity, we will consider unitary RMEs
with broken time reversal symmetry \cite{Mehta}; extension to
other symmetries should be straightforward.
\end{thebibliography}
\end{document}